\documentclass[letterpaper,twocolumn,10pt]{article}
\usepackage{usenix}

\usepackage{graphicx}
\usepackage{latexsym,amssymb,amsmath,amsfonts,mathtools}
\usepackage{siunitx}
\usepackage[hyphens]{url}
\usepackage{hyperref}
\usepackage{breakurl}

\usepackage{listings}
\begin{document}

\date{}

\title{\Large \bf The Serverless Scheduling Problem and NOAH}

\author{
{\rm Manuel Stein}\\
Nokia Bell Labs
} 

\maketitle


\subsection*{Abstract}

The serverless scheduling problem poses a new challenge to Cloud service
platform providers because it is rather a job scheduling problem than a traditional
resource allocation or request load balancing problem.
Traditionally, elastic cloud applications use managed virtual resource
allocation and employ request load balancers to orchestrate the deployment.
With serverless, the provider needs to solve both the load balancing and the
allocation.

This work reviews the current Apache OpenWhisk serverless event load balancing
and a noncooperative game-theoretic load balancing approach for response time
minimization in distributed systems. It is shown by simulation that neither
performs well under high system utilization which inspired a
noncooperative online allocation heuristic that allows tuning the trade-off
between for response time and resource cost of each serverless function.

\section{Introduction}
Serverless is an emerging Cloud Service model that provisions event processing
on demand. In place of traditional VM rental, the providers service dispatching
and execution of events registered with user-provided functions.
This radically changes the deployment model from resource allocation to
on-demand execution. Serverless only charges for the number of dispatched events
and the resource-time covered by the event execution. Amazon\cite{amazon} has
made its serverless platform generally available in Apr'15 and
Google\cite{google}, IBM\cite{IBM} and Microsoft\cite{azurefunctions} have
released similar offerings within a year. As data center efficency had
previously been criticised for comatose and idling
servers~\cite{koomey:2015,koomey:2017}, the shift from a rental model to pay per
use would allow providers to increase server utilization and customers to fit
the cost of a service deployment closer to its actual demand.

This flexibility comes with an overhead to the application execution.
Firstly, serverless applies the stateless worker model, which requires each
function to externalize all context that needs to persist between two
consecutive invocations as the context is not guaranteed to remain in use after
event completion.
Secondly, part of the billed execution time is spent on code initialization and
data access at the start of processing, which is small compared to VMs, but can
still be significant in comparison to the total runtime of a function
(100ms-5min).
Thirdly, the setup time contributes a substantial overhead to the response time
(up to 1-2s if it scales to an additional container).
The service model devolves responsibility to solve this scheduling challenge to the
provider. Traditionally, cloud-native application orchestration would require
development of application-specific load balancers and engineering to find a
suitable trade-off between response time guarantees, data locality and proactive
resource allocation. The serverless provider is entrusted with scaling and would
ideally adapt to the demand. 

Serverless is supposed to ease the
orchestration and operation automation challenge as it finds the application
decomposed into a functional (i.e. event-, or data-driven) set of non-blocking 
stateless executions that are structurally similar tasks which can be scaled
independently. The provider challenge is to find a demand scheduling that fits
all applications' performance requirements.
The decomposition into stateless, short-lived executions that perceive increased
(context) data access latency and initialization time calls for locality-aware
scheduling to maximize cache hit rates. Affinity scheduling is common in high
throughput computing. It can be found in VM placement policies for network
colocation in NFV Clouds and at task level in microbatching platforms.
But locality poses a challenge when providers aim for high resource utilization
whereas customers strive for minimum response times.
The balance between maximizing resource utilization and minimizing response time
is a multi-objective problem. This work reviews state-of-the-art demand
scheduling approaches and introduces a noncooperative online allocation
heuristic (NOAH) as a parameterizable approach to configure the serverless
scaling behaviour.

Section \ref{sec:problem} classifies the serverless scheduling problem and
describes the optimization problem, followed by a review of the current
OpenWhisk scheduling heuristic and a game-theoretic approach to response time
minimization. Section \ref{sec:noah} introduces the new allocation heuristic
NOAH to solve the serverless scheduling problem. Section \ref{sec:simulation}
describes the simulation used to evaluate the presented approaches. Section
\ref{sec:related} discusses related work and section \ref{sec:concl} concludes
the findings.

\section{Related Work}\label{sec:related}

Amazon Lambda has recently introduced a limit to concurrent executions per
functions, which would contain the number of started instances but neither
considers response time nor scaling behaviour.
Microsoft Azure, IBM Cloud Functions and Google Cloud Functions provide a quota
of concurrent executions per user or namespace but not for individual functions.

McGrath and Brenner~\cite{mcgrath:design} design a classful system that lets 
instances poll for work. NOAH makes no specific assumptions about the messaging 
and can be implemented with polling.
CaaS-based serverless platforms such as Kubeless~\cite{kubeless},
Fission~\cite{fission}, OpenFaaS~\cite{openfaas} adhere to container resource
allocation, scale by resource monitoring and use hash-based reverse proxy load
balancing as best practices of cloud-native platforms.

Multiple works have inspired NOAH, i.e.
smoothing of online virtual resource allocation to save reallocation
cost\cite{jiao:2017}, game-theoretic optimally controlled load
balancing\cite{grosu:2005,grosu:2008}, multi-objective game-theoretic job
scheduling\cite{duan:2014}, distributed, classful optimal control demand
reallocation\cite{widjaja:2013} and integration of queuing theory with
scheduling\cite{terekhov:2014}.




\section{The Serverless Scheduling Problem} \label{sec:problem}

In serverless platforms, the scheduling subsystem dispatches events from their
occurrence (at a gateway or within the platform) to the designated resource that
performs the function execution. The demand follows an open arrival process, is
distributed across the platform and has multiple users per function. All tasks
have a similar structure and are, in principle, processes with moldable
execution time in an OS-sharing model. The problem can be classified as
distributed job scheduling applying the taxonomy by Lopez and Menasc\'e~
\cite{lopez:2016}, which positions it quite uniquely among popular research.

Traditional Cloud VM scaling is sluggish and separates the allocation control
loop from request load balancing. On the contrary, emerging serverless
platforms decide upon every event whether they can reuse an available instance 
(warm container) or simply launch a new one (cold container).
As a result, bursts of events cause the launch of many parallel instances as the
heuristic provides no control over the scaling behaviour and is only constrained
by resource quota.

When an event is dispatched to a site, the container to host the instance may
need initialization (cold start), it may miss only the function code
(pre-warmed) or have already run the function (warm). The first cold start of
its type may require the host to load custom function code (and library
dependencies), which may require network transfer or disk access. Other
techniques may provide fast copies of an instance (e.g. forking) or its
container context. The total \emph{setup time} of an instance may vary significantly w.r.t.
the function execution time under dynamic workloads.

Execution time starts with the handover of the event message to the function
instance. Depending on its content, named data are accessed from the platform's
data store, e.g. remote or host-local replica, residing on disk or in memory.
Although the function runtime is typically memory-heavy compared to the data
transfers in question, data access and concurrent modification can
cause \emph{access latency} that varies based on the location. Content population of 
cold caches incurs additional execution overhead.

The serverless provider bears the resource cost of the \emph{setup time} and the
customer pays for \emph{access latency}, so both the platform and the function
share a common contractual objective to reduce the makespan of executions while 
keeping the resource use to a required minimum.
Each function competes against others for colocation to reduce access latencies.

\subsection{Noncooperative Load Balancing} \label{sec:nlb}
Grosu and Chronopoulos \cite{grosu:2005} have designed a non-cooperative game
approach to balance multi-user flow allocations for optimal response time in a
distributed server environment. Each user (e.g. distributed controller) collects
information on the service time and allocations at every host to calculate an
optimal split of its own perceived arrival rate in response to the other
players' allocations.

The noncooperative approach yields Nash Bargaining Solutions for a set
of heterogenous M/M/1-type servers (a cooperative version is
described in \cite{grosu:2008} that creates Pareto-optimal solutions).
It is a queuing-analytic, distributed, multi-user solution but remains 
classless. To apply it to the OpenWhisk controller
architecture, each function is made a player in the noncooperative game to compete in the allocation for minimum response time.

\subsection{OpenWhisk Heuristic} \label{sec:ow}
The Apache OpenWhisk serverless platform implements event scheduling in a
distributed controller architecture that shares common host state information,
including the number of concurrently active invocations on each host.
A function name's hash $h$ is used to identify a starting point in the pool of
sites $I$.
If the number of active events at the target exceeds a threshold ($\alpha$), the
controller tries the next host in pseudo-random order using pairwise coprime
generators as the following pseudo-code illustrates.

\begin{lstlisting}
select_host(h=$\textrm{hash}$, I=[]$\textrm{sites}$) {
  lvl=$\alpha$=16, G=[]
  for(i$= 0,\ldots,|I|$) { $\textrm{// create generators}$
    if($\forall g \in \text{G}, gcd(i,g) == 1$)
      G=G $\cup$ i
  }
  g=G[$\text{h} \pmod{|\text{G}|}$] $\textrm{// select generator}$
  while(lvl $\leq 3\alpha$){ $\textrm{// try up to 3 times busy threshold}$
    for(k$= 0,\ldots,|I|$) {
      x = I[$\text{h} + \text{k}*\text{g} \pmod{|\text{I}|}$]
      if($\textrm{events at}(\text{x})$ < lvl)
        return x
    }
    lvl += $\alpha$
  }
  return $\textrm{random site}$
}
\end{lstlisting}

If no host is below the busy level, the threshold is increased and the
pool is progressed again. Eventually, a random site would be picked.
This compelling first-fit, hash-based load balancing heuristic solves
statistically the locality, distribution and overflow of function allocations
across hosts,
as it ensures that all resource slots are in use before the system starts to
oversubscribe its resources.
However, the host progression does not consider the instance setup time when
scaling out a function type to a host that hasn't previously been employed. For
some events, it may be beneficial to enqueue rather than scale out.



\begin{figure*}[t]
    \minipage{0.33\textwidth}%
        \includegraphics[width=\linewidth]{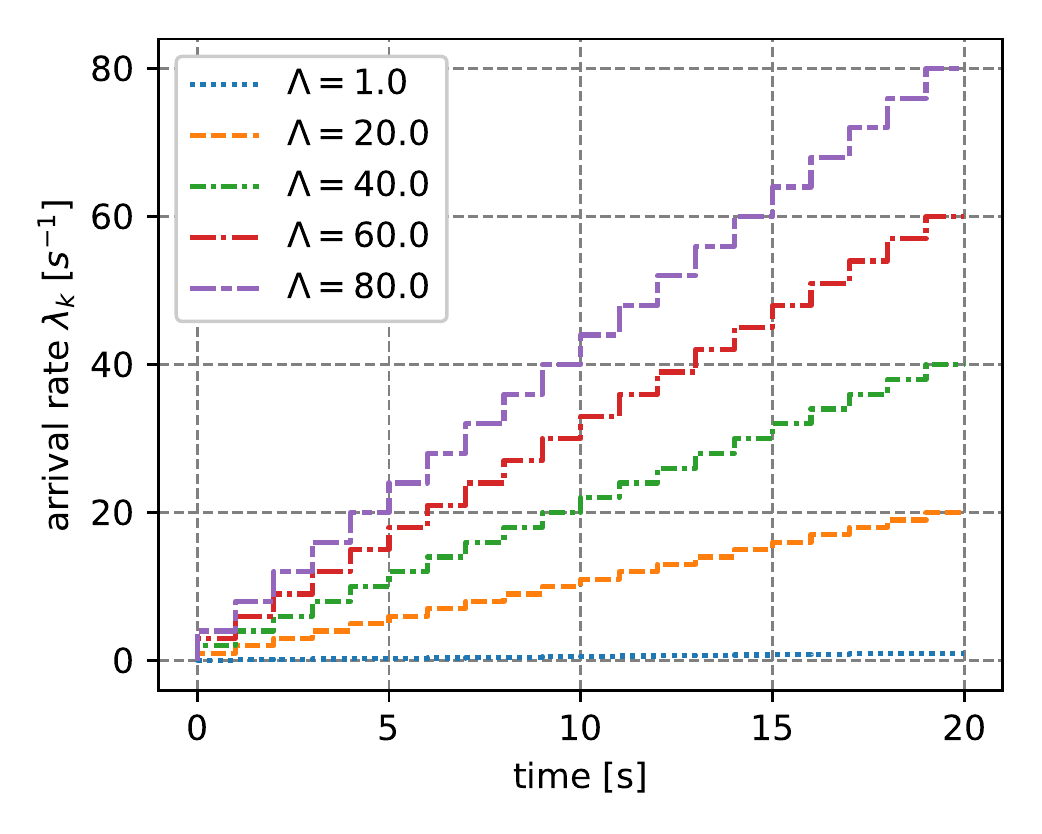}
	    \caption{Workload $\lambda_k(t) = \lceil t \rceil * \Lambda$}
	    \label{fig:explain}
    \endminipage
    \minipage{0.33\textwidth}%
	    \includegraphics[width=\columnwidth]{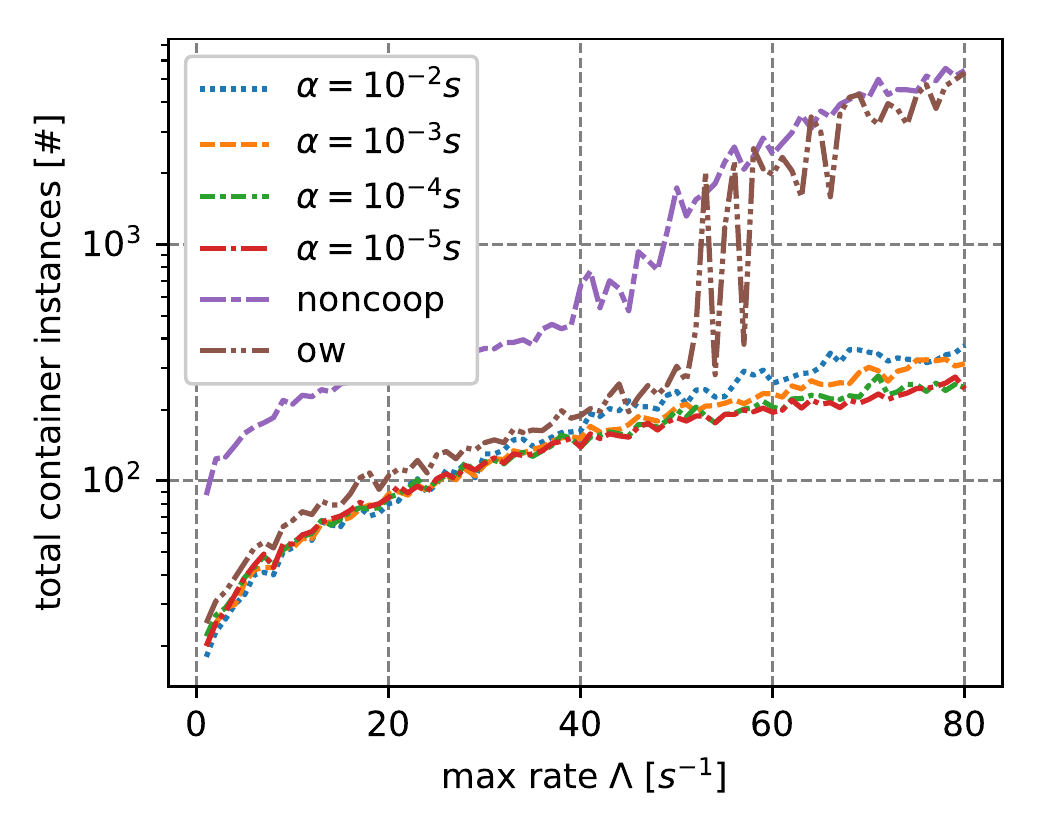}
	    \caption{Cold starts}
	    \label{fig:container}
    \endminipage
    \minipage{0.33\textwidth}%
        \centering
	    \includegraphics[width=\columnwidth]{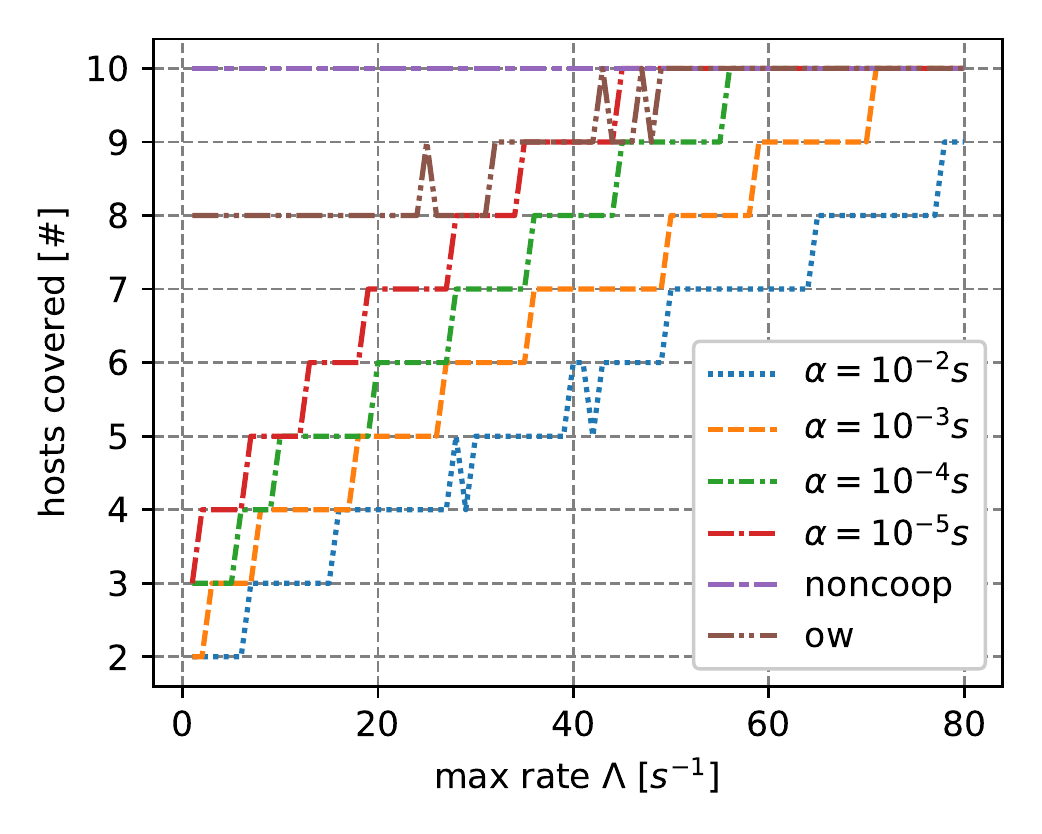}
	    \caption{Hosts employed}
	    \label{fig:worker}
    \endminipage
\end{figure*}


\section{NOAH}\label{sec:noah}

The two approaches of hash-based and
queue-analytical minimum-response time balancing presented in section \ref{sec:problem} face high churn rates of
instances when demand increases: noncooperative load balancing tries to
statistically distribute events across hosts, so different function types evict
each other's instances upon changes in allocation, resulting in an increased
occurrence of instance setup times. OpenWhisk uses pseudo-random progresison to
find a free spot where it also would evict an idling container.

This has inspired the design of a noncooperative online allocation heuristic
(NOAH), which uses both an analytic allocation and a minimum
completion time heuristic to design a configurable solution to the serverless
scheduling problem. The heuristic estimates for each function type the required
number of instances to keep expected request waiting time below a chosen
threshold. The resulting allocations are then placed with sites, but no actual
instances are spawned by an allocation.
Instead, each site manages the instance pool to handle arriving requests,
deciding autonomously whether to enqueue a request or spawn a new instance.
Event dispatching tries to schedule events to idling instances for minimum
completion time and otherwise balances requests according to the allocations of
the function.

\noindent\textbf{Allocation.} Each function uses an analytical model to contain the
expected average waiting time that an event would face in the system. Estimating
the arrival rate $\hat\lambda_k$ for a function type $k$ and the current average
service time of function executions $\hat\mu_k$, the number of required (warm)
instances $c_k$ can be chosen, such that the expected waiting time of queuing
remains below threshold $\alpha$.


For example, assuming an M/M/c model, i.e. a Poisson arrival process and
exponentially distributed service times, the allocation should at least satisfy
the stability condition: $\frac{\lambda_k}{c_k\mu_k} < 1 \implies c >
\frac{\mu_k}{\lambda_k}$. To contain the probability that an arriving event
finds all instances active, the Erlang C formula can be used to find $c_k$ such
that $\frac{C(c_k,\frac{\lambda_k}{\mu_k})}{c_k\mu_k - \lambda_k} < \alpha$.

Having identified the required number of allocations $c_k$ for class $k$ events,
the system needs to place allocations $c^i_k, \sum_{i}{c^i_k} = c_k$ to sites
$I$ with a maximum number of allocations per site $c^i$ such that
$\sum_{k}{c^i_k} <= c^i$.

\noindent\textbf{Dispatching.} Following the OpenWhisk example of distributed
host state information, the controller can obtain if any host has an idling
instance for the given function type to schedule the event. If there is no
idling instance, the event is dispatched to the host with the lowest ratio of
active instances over allocations.

\noindent\textbf{Execution.} Each site measures separate setup and execution
time of a function type starting from shared experience values. If an arriving
event finds an empty queue and an idling instance, it is immediately scheduled.
If the maximum number of actively processing events is reached, events are
queued. But, if the host can chose between queuing the event or starting a new
container, it checks whether the estimated queue waiting time (given the
estimated completion of concurrent events) would exceed the estimated setup
time, in which case it would opt to launch a new container.



%

\subsection{Allocation Heuristic}

While the allocation control loop of every event class contains the average
waiting time, the concertated allocation management needs to select hosts when a
class wants to scale-out or scale-in.

The allocation control for each event class constitutes a player in a
noncooperative game for the location of the allocations. Depending on the effect
of \emph{access latency} on the response time of events, classes benefit
differently from instance colocation, e.g. when synchronizing shared context. 
When it scales out, allocation management tries to colocate allocations. This
heuristic reduces the setup time overhead to transfer dependencies and populate
caches to occasions when it is inevitable.
When a function scales in, the host with the least allocations is reduced.
This keeps the bulk of events colocated and decreases both expected waiting time
of an arriving event and may be used to reduce data synchronization latency.
Classes with a shared data context have a stronger incentive to aggregate
allocations, so coalitions to swap allocations (tit-for-tat) between classes are
possible and defragmentation strategies are possible, but currently not implemented.

\section{Simulation}\label{sec:simulation}

To evaluate the approach, a serverless simulation tool was developed using SimPy
(v3.0.1) to mimic the Apache OpenWhisk architecture which dispatches events
via a controller to invokers that manage local container pools with a maximum
size each.

The simulation execution uses processor sharing and memory caching. Executions
can be purely processing or read/write accesses to named data items which
requires replication from the least loaded copy to local memory. Transfer incurs
execution at both ends and honors solid state disk, memory and network speed.


Work-conserving scheduling has been verified using several queuing models. Data
transfer, access and eviction has been verified with a set of tests that assert 
precedence. Ultimately, the simulation tool has been verified running designed 
workloads on both OpenWhisk and the simulation tool. Simulation timing of container 
setup and function initialization has been directly obtained from OpenWhisk 
invoker logs.

Noncooperative load balancing 
is implemented as an alternative controller (\emph{noncoop}). For
NOAH
, modifications to both the controller as well
as the invoker simulation were necessary.




\begin{figure*}[t]
    \minipage{0.5\textwidth}%
		\centering
	    \includegraphics[width=0.85\linewidth]{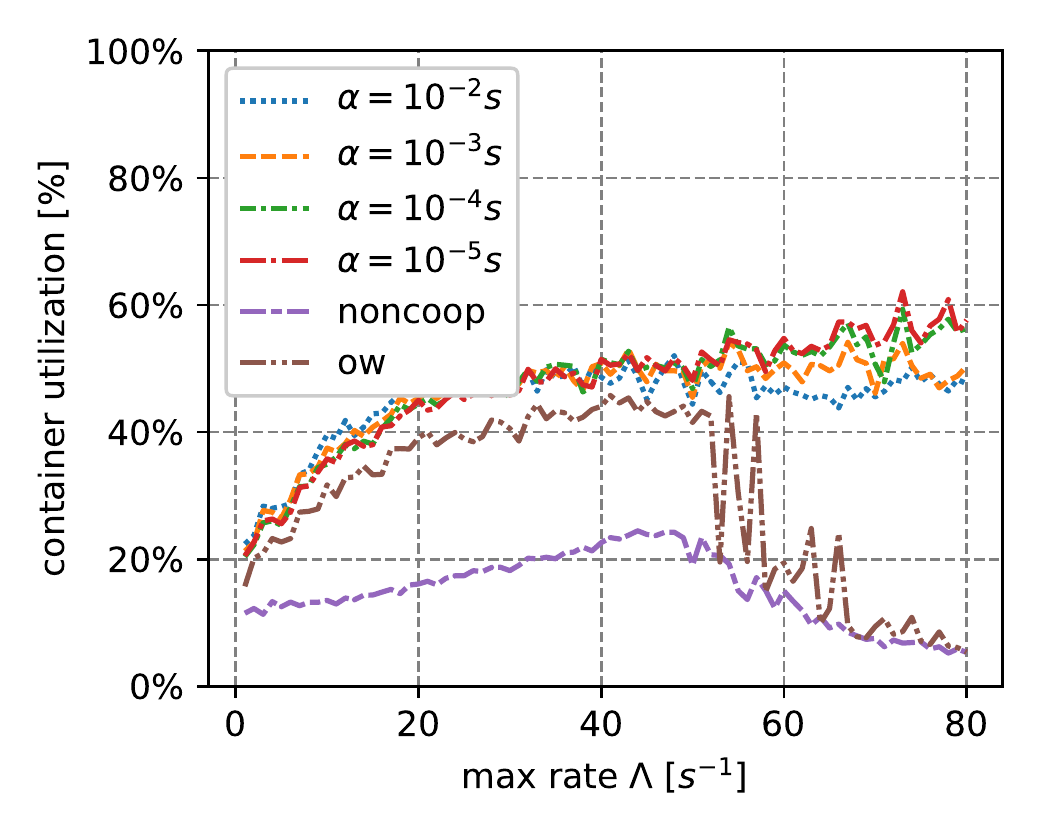}
    	\caption{Container utilization}
	    \label{fig:utilization}
    \endminipage
    \minipage{0.5\textwidth}%
	    \centering
		\includegraphics[width=0.85\linewidth]{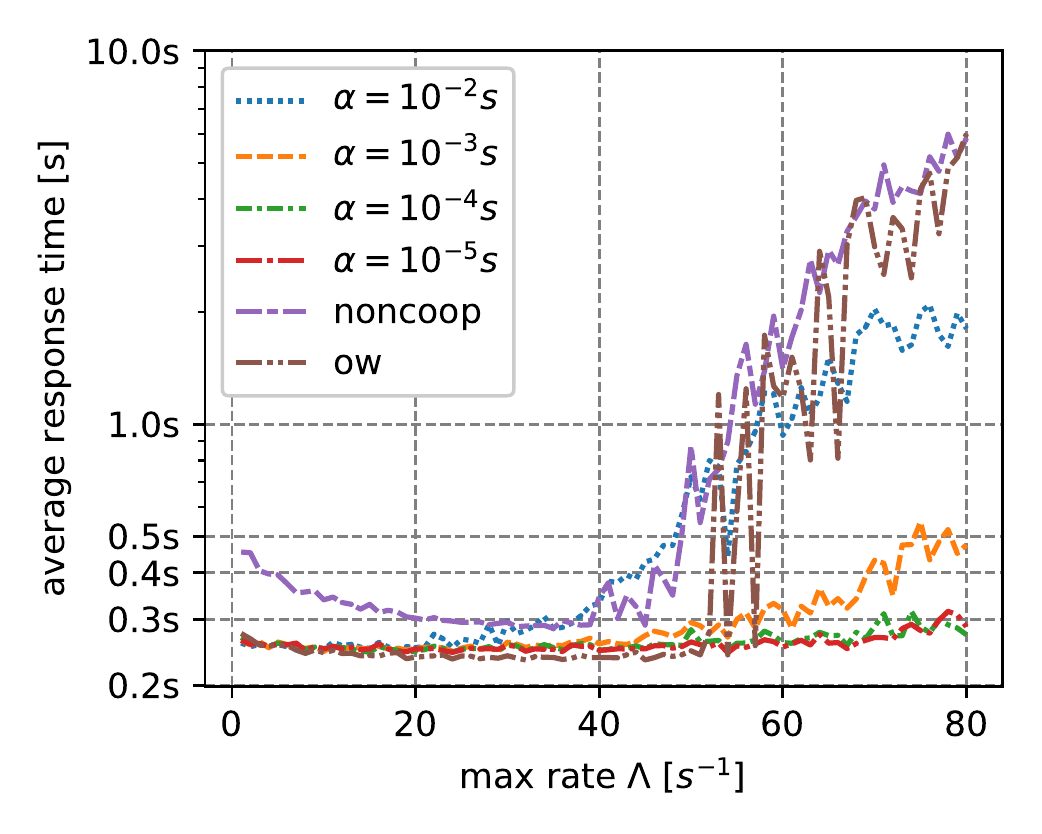}
	    \caption{Average response time}
    	\label{fig:response_time}
    \endminipage
\end{figure*}

\subsection{Evaluation}

The simulated system has ten homogenous hosts with 16 cores each, 711MB/s disk
speed (SSD), 1,135MB/s network speed (10GbE) and 12.8GB/s memory speed with
sufficient capacities each.
The evaluation scenario simulates ten function types, each with an independent
Poisson arrival process. To measure the scaling behaviour, a series of
simulations has been run with different slopes and load maxima $\Lambda$.
The arrival rate of function types equally increases as depicted in
figure~\ref{fig:explain} for different simulations. In each simulation, the
interarrival rate to each function scales within \SI{20}{\second} from zero to the
maximum arrival rate $\Lambda$ and then stops. The simulation is run until all
messages have been processed.
Each function is single-threaded and has an ideal execution time of \SI{200}{\milli\second} -
unless processor sharing slows execution, thus the upper
bound system arrival rate is $10 hosts * 16 cores * 5\frac{msg}{sec} = 800$.
Simulations are run with 10 functions and $\Lambda =
\{1,2,\ldots,80\}\frac{msg}{sec}$ each. Due to container setup (\SI{500}{\milli\second}), message
transfer and suboptimal balance, the approaches start struggling at about 50-60
messages per second per function.

The remainder discusses the performance of NOAH with waiting time thresholds $\alpha$ set to \SI{10}{\milli\second}, \SI{1}{\milli\second},
\SI{100}{\micro\second} and \SI{10}{\micro\second}, against default OpenWhisk \emph{ow} and noncooperative load balancing \emph{noncoop}.

\noindent\textbf{Cold starts.} OpenWhisk invokers manage a configurable maximum
of 32 containers. The default busy threshold causes the controller to scale-out
at 16 concurrently processing events. Only 16 containers can be active in
parallel to avoid processor sharing and context switching. The gap prevents a
host from evicting containers every time a function does not run on its primary
host target. The default holding time for an unused container is \SI{5}{\minute}
unless the container needs to be evicted earlier. The more often controllers use
overflow progression, the more eviction occurs and in consequence, the more cold
starts occur.
Noncooperative load balancing \emph{noncoop} almost equally balances each
function's load across sites, which likewise evicts idling containers to keep a
maximum of 32 instances. With \emph{noncoop}, flows spread across the whole pool
for minimum latency, which causes at least one instance of a function type at
every server and a high container churn under high load.
NOAH invokers limit the number of concurrent events and the total number of
containers separately. The total number of instances (active and inactive) is 
bound by the number of allocations and the host memory capacity.

Figure~\ref{fig:container} shows the total number of containers started in each
simulation. At $\Lambda = 50$, the load causes OpenWhisk to resort for some
events to random site selection and therefore, eviction. NOAH can maintain
stable increase of cold starts proportional to the workload increase.

\noindent\textbf{Hosts covered} by request processing counts only those sites 
that have been employed to process at least one event. Load balancing approaches 
\emph{ow} and \emph{noncoop} always try to utilize the entire pool. OpenWhisk may in
rare cases see hosts that would only be used for excess load, which explains why
low rates keep 1-2 hosts unused.
NOAH's allocation tuning covers only the required minimum set of hosts and scales out when necessary.
NOAH allows unused hosts (e.g. VMs) to be released entirely, reducing resource
cost to the required minimum. Tuning $\alpha$ to low expected waiting
times requires more hosts whereas an allowance for longer waiting times covers
less hosts in total.

\noindent\textbf{Average response time} is best 
when instances can be reused (fig.~\ref{fig:response_time}). OpenWhisk maintains the best response times
under moderate load, but suffers tragically under high utilization, 
because it can not contain the container
churn. NOAH's allocation tuning of the waiting time using the Erlang-C formula
shows clear increments between tuning for low response times ($\alpha$=\SI{10}{\micro\second}) to
high response times ($\alpha$=\SI{10}{\milli\second}) antiproportional to the resource coverage.

\noindent\textbf{Container utilization} (fig.~\ref{fig:utilization}) is the
ratio between the time a container actively processes an event (execution time)
and the total time that it remains in the system (including setup and idle
overhead). Regardless the container timeout (which would add \SI{5}{\minute} idle time to
the \SI{20}{\second} experiment), only the time from a container's creation until the last
processing of a message is considered here as the total container life time.
Overall, NOAH yields higher container utilization. For \emph{ow} and
\emph{noncoop}, increasing eviction reduces utilization.
Under high load, the allowance for a longer queuing delay ($\alpha$=\SI{10}{\milli\second})
conversely shows lower utilization than stringent requirements ($\alpha$=\SI{10}{\micro\second}).
Notice also, that a ($\alpha$=\SI{10}{\milli\second}) covers less hosts than ($\alpha$=\SI{10}{\micro\second})
(cmp. fig.~\ref{fig:worker}), which leads to higher churn of instances as can also
be seen in the number of container creations (cmp. fig.~\ref{fig:container}).

\section{Conclusion}\label{sec:concl}

For serverless event scheduling, NOAH shows several benefits over the current OpenWhisk
controller\cite{openwhisk} and the game-theoretic load balancing
approach\cite{grosu:2005}.
By virtually allocating demand, NOAH can contain instance churn 
and remains responsive under high system utilization. 
Furthermore, it limits execution to required sites which allows scaling of 
the resource pool and provides means to balance the trade-off between resource 
cost and response time by tuning for allowed queue waiting times of individual functions.




{\footnotesize \bibliographystyle{acm}
\bibliography{ms}}


\end{document}